# Subwavelength imaging of sparse broadband sources surrounded by an open disordered medium from a *single* antenna

Lianlin Li,[1,*] Fang Li,[2] and Tie Jun Cui[3]

[1]*School of EECS, Peking University, Beijing, 100871, China*
[2]*Institute of Electronics, Chinese Academy of Sciences, Beijing, 100080, China*
[3]*State Key Laboratory of Millimeter Waves, Southeast University, Nanjing, 210096, China*
[*]*lianlin.li@pku.edu.cn*

In this letter we study the subwavelength imaging of sparse broadband sources inside a disordered medium by processing the data acquired by a *single* antenna. A mathematical model has been developed for solving such problem based on the idea of sparse reconstruction. We show that the strongly disordered medium can serves as an efficient apparatus for compressive measurement, which shifts the complexity of devising compressive sensing (CS) hardware from the design, fabrication and electronic control. The proposed method and associated results can find applications in several imaging disciplines, such as optics, THz, RF or ultrasound imaging.

*PCAS Codes:*

Over past years, a family of strategies has gained researchers' intensive attentions, which explores the multiply scattering of wavefields in a strongly disordered medium (or scattering medium) to enhance the quality of focusing and imaging in various applications, such as acoustics, seismology, microwave, optics, etc [1-5]. For these schemes, an array of detectors is mandatorily required to obtain an imaging with high spatial resolution, for instance, the well-known time-reversal mirror (TRM in short). Recently, in the context of broadband time reversal, Pierrat et al investigated theoretically the feasibility of subwavelength focusing in an open disordered medium with a *single* antenna [1]. The goal of this letter is to achieve the subwavelength imaging of sparse broadband sources surrounded by the strongly disordered medium.

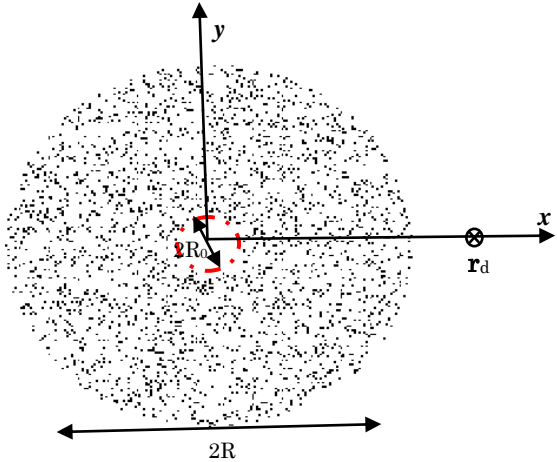

FIG. 1 (Color online) System composed of a 2D cluster of nonabsorbing cylindrical scatters randomly placed inside a cylinder of radius R centered at (0, 0) where the broadband sources under investigation are located at the exclusion circle of radius $R_0$. Detection with a single point sensor is located at $\mathbf{r}_d$ in the far field. The radius of cylindrical scatters is 22nm, relative permittivity $\varepsilon_r =8$, and volume fraction 10%.

Mathematically, this problem is conventionally reduced into solving a linear inverse problem. Such problem is usually ill-posed because of inadequate measurements relative to the number of unknowns, so that its solution is severely non-unique. The recent theory of compressive sensing (CS in short) demonstrates that the reduced significantly number of measurements, accomplished by acquiring an incomplete set of pseudo random projections, are sufficient to recover a sparse or compressible signal under test by employing a tractable nonlinear program [e.g., 6, 7]. Now, a natural question arise up: is it possible to obtain the subwavelength imaging of sparse broadband sources by exploring the CS related results? The answer is encouraging. As demonstrated below, a strongly disordered medium can serve as a natural candidate for compressive measurements related to the subwavelength imaging, since each measurement gathers information from all parts of the imaged object in a controlled but pseudo-random fashion owing to the multiple scattering effect (or multiple propagation paths) in a disordered medium.

Similar to the setup adopted in Ref. [1], we consider a two-dimensional (2D) cluster of nonabsorbing cylindrical scatters randomly distributed inside a cylindrical region of radius R=2.1μm (see Fig. 1) in this letter. The radius of cylindrical scatters is 22nm, and a minimum distance 5nm is forced between scatters to avoid overlapping. For numerical simulation, the range of operational wavelength is from 538nm to 838nm with a separation of 1nm. These cylindrical scatters are aligned infinitely along z-axis, and the probed sources are polarized along the scatters (TE mode), such that the electromagnetic problem is scalar. These sources under investigation lie on the circle of radius $R_0$, centered at the origin. Note that the value of $R_0$=344nm is chosen so as to preserve adequate near-filed interaction between probed sources and scatters, as suggested in Ref. [1]. A single point antenna for collecting data lies in the far field, outside of the system, at position $\mathbf{r}_d$=(0, 6.8)μm.

First, we apply the coupled-dipole method [8] to derive the Green's function for the whole system $G(\mathbf{r}_d, \mathbf{r}_\theta; \omega)$, which relates the electric field at the position of antenna $\mathbf{r}_d$ to a given point source at $\mathbf{r}_\theta$=($R_0\cos\theta$, $R_0\sin\theta$), $\theta \in [0°, 360°)$. We detail this procedure in appendix. Figure 2 (a) and (b), respectively, shows the normalized amplitude and phase of the Green's function $G(\mathbf{r}_d, \mathbf{r}_\theta; \omega)$ associated with the one configuration as shown in Fig. 1. In these figures, the x-axis denotes the value of $\theta$ in degree, while the y-axis is for the operational wavelength in nm.

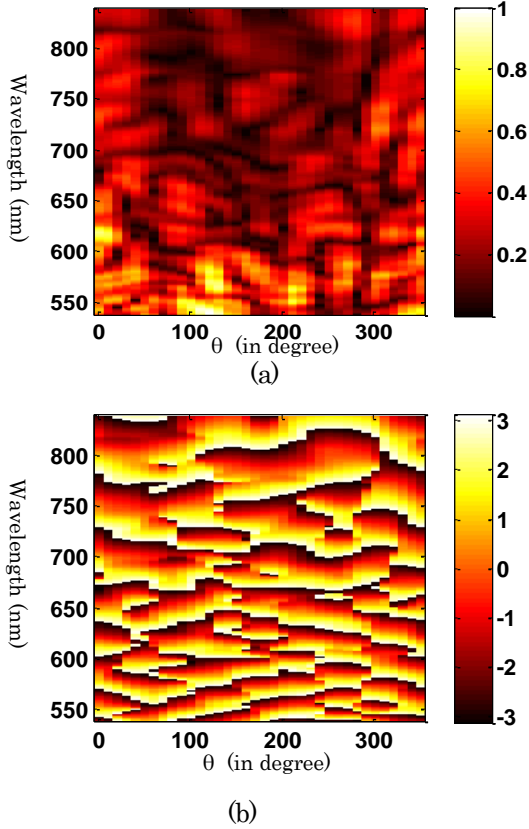

(a)

(b)

FIG. 2. (Color online) The normalized amplitude and phase of the Green's function $G(\mathbf{r}_d, \mathbf{r}_\theta; \omega)$ associated with the one configuration as shown in Fig. 1. In this set of figures, the x-axis denotes the value of θ in degree, while the y-axis is the operation wavelength in nm.

Now, the problem of finding the broadband sources $p(\omega, \theta)$ can be casted into solving following minimum least-square problem, i.e.,

$$\min_{p(\theta;\omega)} \int [y(\omega) - \sum_\theta G(\mathbf{r}_d, \mathbf{r}_\theta; \omega)p(\theta; \omega)]^2 d\omega \quad (1)$$

where $y(\omega)$ represents the data at the angular frequency of ω acquired by the sensor at $\mathbf{r}_d$ and the discrete values of θ varies from 0° to 360° in the separation of 10°. Obviously, Eq. (1) is ill-posed since the measurements available are seriously inadequate relative to the number of unknowns. To tackle this issue, Eq. (1) is regularized through the mixed norm of $p(\omega, \theta)$, as a result, one has

$$\min_{p(\theta;\omega)} \{\int [y(\omega) - \sum_\theta G(\mathbf{r}_d, \mathbf{r}_\theta; \omega)p(\theta; \omega)]^2 d\omega \quad (2)$$

$$+\gamma \sum_\theta \frac{|p(\theta;\omega)|^2}{\int |p(\theta;\omega)|d\omega}\}$$

where γ is a factor used to balance the data fidelity and regularization term. In Eq. (2), we exploit the fact that $p(\omega,\theta)$ shares the common support for a given θ, which relieves efficiently the ill-posedness of Eq. (1). Considering further that the broadband sources under test exhibit some coherence among them, we express $p(\omega,\theta)$ in a reduced dimension, i.e.,

$$p(\omega,\theta) = \sum_{m=1}^{M} p_m(\theta)\varphi_m(\omega) \quad (3)$$

where $\{\varphi_m(\omega), m=1,2,…,M\}$ is a set of known basis, or atoms [6] for describing $p(\omega,\theta)$. Here, M is very small relative to the dimension of ω in the discrete form. Hence, Eq. (2) can be modified as

$$\min_{p_m(\theta)}\{\int[y(\omega) - \sum_\theta \sum_{m=1}^{M} G(\mathbf{r}_d, \mathbf{r}_\theta; \omega) p_m(\theta)\varphi_m(\omega)]^2 d\omega + \gamma \sum_\theta \frac{|p_m(\theta)|^2}{\sum_{m=1}^{M}|p_m(\theta)|}\} \quad (4)$$

Note that if M=1, Eq. (4) is reduced into the model used in Ref. [1]. To solve Eq.(4), the well-developed iteratively reweighted algorithm is performed [6, 9].

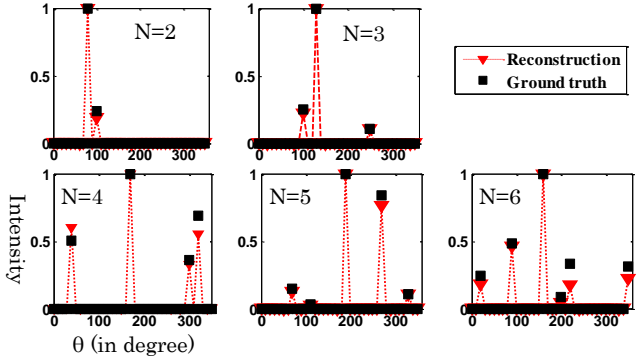

FIG. 3. (Color online) The normalized intensities reconstructed by proposed methodology for varying number of broadband sources N. In simulations, the Fourier basis with reduced dimension of M=5 is adopted. The x-axis denotes the value of θ in degree, while the y-axis is for the normalized intensity.

Figs. 3 display the normalized intensities reconstructed by solving Eq. (4), represented by $I(\theta) = \int |p(\theta;\omega)|^2 d\omega$, for varying number of probed broadband dipoles, where the ground truth is denoted by black filled squares, and the reconstruction is denoted in red-filled triangles. In our simulations, the minimum separation between two adjacent dipoles is set to be $10^0$, corresponding to 0.078λ of the minimum resolved distance. For the sake of simplicity, the Fourier-basis $\varphi_m(\omega) = e^{i\omega\frac{(m-1)}{B}}$ is adopted here. Moreover, M=5 is assumed in Fig. 3. From this set of figures one can see that multiple broadband dipoles can be easily distinguished in the subwavelength spatial scale.

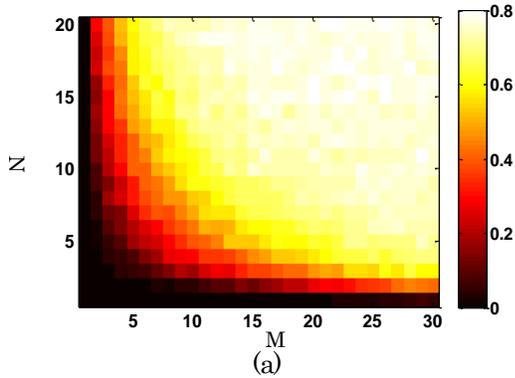

(a)

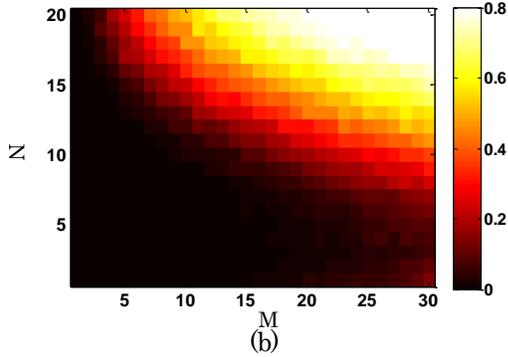

FIG. 4. (Color online) The behaviors of obtained MSE as a function of the number of basis, M, and the number of broadband sources, N. (a) the Fourier basis, (b) the Gaussian basis. (The MATLAB code for reproducing Figs. 4 can be freely achieved by sending a request email to lianlin.li@pku.edu.cn)

Now we will examine the effect of $M$ on the accuracy of reconstructed intensities, and investigate the capability of proposed methodology for locating the number of broadband dipoles. Fig. 4(a) shows the normalized MSEs of reconstructed intensities as a function of the number of Fourier basis M, and the number of probed sources N, which is obtained and averaged over 50 Monte-Carlo independent trials. The normalized MSE is defined as

$$\text{MSE} = \frac{\sqrt{\sum_\theta [\int |\hat{p}(\theta;\omega)|^2 d\omega - \int |p_0(\theta;\omega)|^2 d\omega]^2}}{\sqrt{\sum_\theta [\int |p_0(\theta;\omega)|^2 d\omega]^2}}$$

(5)

where $I = \int |\hat{p}(\theta;\omega)|^2 d\omega$ is the reconstructed intensity, and $I = \int |p_0(\theta;\omega)|^2 d\omega$ corresponds to the truth. From above results, one conclusion can be immediately deduced that the higher the value of M is, the worse the successful probability is. This makes sense since the amount of unknowns is proportional to M. In addition, to investigate the effect of $\{\varphi_m(\omega), m=1,2,\ldots,M\}$ on the quality of retrieved sources by proposed method, we consider a special case that the entries of $\{\varphi_m(\omega), m=1,2,\ldots,M\}$ draw from i.i.d. Gaussian random numbers with zero mean and unit variance. For notational convenience, this basis is termed as the Gaussian basis. For the purpose of comparison, associated results are provided in Fig. 4 (b), which confirms above conclusion. Moreover, the reconstruction with such random Gaussian basis is better than that with Fourier basis. This comparison reveals to some extent that it is easier to locate the random sources than to determine the locations of well-behavioral sources. It is worth remarking that despite a relatively big error for the case of Fourier basis, the locations of probed broadband sources still can be clearly identified for moderate values of M and N. To see it, the reconstructed intensities are compared in Fig. 5 for the specific choice of M =10 and N= 8. From this figure, we can see that despite the MSE being relatively big for the Fourier basis, the locations of 10 broadband sources can also be easily determined.

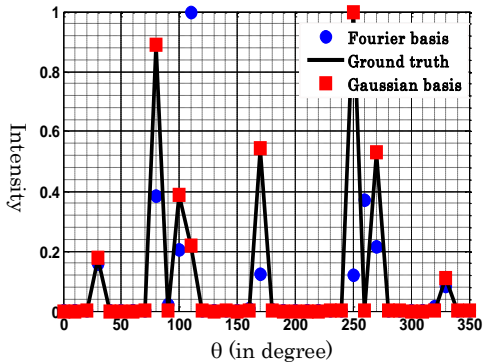

FIG. 5. (Color online) The normalized intensities reconstructed by proposed methodology for the case of M=10 and N=7. The ground truth is denoted by the black line, the reconstruction with Fourier basis is denoted by blue circles, and the reconstruction with Gaussian basis is denoted by red squares. The x-axis denotes the value of θ in degree, while the y-axis is the normalized intensity.

To look further insight into above conclusions, we expand the first term of Eq. (5) in the form of linear equations, namely,

$$\begin{bmatrix} y(\omega_1) \\ \vdots \\ y(\omega_k) \\ \vdots \\ y(\omega_K) \end{bmatrix} = \begin{bmatrix} G(\mathbf{r}_d,\mathbf{r}_{\theta_1};\omega_1)\boldsymbol{\varphi}_1 & \cdots & G(\mathbf{r}_d,\mathbf{r}_{\theta_n};\omega_1)\boldsymbol{\varphi}_1 & \cdots & G(\mathbf{r}_d,\mathbf{r}_{\theta_N};\omega_1)\boldsymbol{\varphi}_1 \\ \vdots & \vdots & \vdots & \vdots & \vdots \\ G(\mathbf{r}_d,\mathbf{r}_{\theta_1};\omega_k)\boldsymbol{\varphi}_k & \cdots & G(\mathbf{r}_d,\mathbf{r}_{\theta_n};\omega_k)\boldsymbol{\varphi}_k & \cdots & G(\mathbf{r}_d,\mathbf{r}_{\theta_N};\omega_k)\boldsymbol{\varphi}_k \\ \vdots & \vdots & \vdots & \vdots & \vdots \\ G(\mathbf{r}_d,\mathbf{r}_{\theta_1};\omega_K)\boldsymbol{\varphi}_K & \cdots & G(\mathbf{r}_d,\mathbf{r}_{\theta_n};\omega_K)\boldsymbol{\varphi}_K & \cdots & G(\mathbf{r}_d,\mathbf{r}_{\theta_N};\omega_K)\boldsymbol{\varphi}_K \end{bmatrix} \begin{bmatrix} \mathbf{p}_1 \\ \vdots \\ \mathbf{p}_n \\ \vdots \\ \mathbf{p}_N \end{bmatrix}$$

(6)

where K and N denotes the number of observation frequencies and discretized θ for numerical computations, $\boldsymbol{\varphi}_k = [\varphi_1(\omega_k), \varphi_2(\omega_k), ..., \varphi_M(\omega_k)]$, and $\mathbf{p}_n = [p_1(\theta_n), p_2(\theta_n), ..., p_M(\theta_n)]^T$. Note here N is different from previous one. Among all the features that were proposed to characterize a matrix as a good candidate for CS, the coherence plays a special role because of its easy-implementation [6]. For this consideration, Fig. 6 compares the coherences for the cases of the Fourier basis and Gaussian basis. By CS theories [6], one can confirm above conclusion that it is easier to locate the random sources than to determine locations of well-behavioral sources. Anyway, our studies suggest that the disorder medium can be a good candidate in a CS setup for subwavelength imaging of sparse broadband sources.

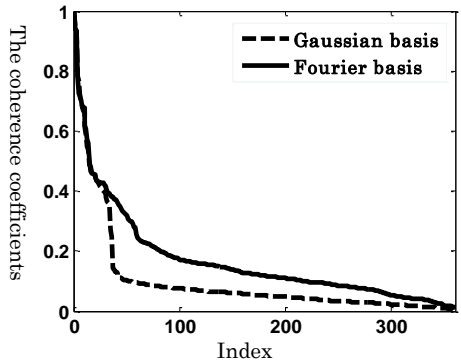

FIG.6 (Color online) The coherence coefficients of mapping matrix in Eq. (8) for Gaussian basis (dashed line) and Fourier basis (sold line).

In summary, we develop a novel scheme of the imaging of sparse broadband sources in a subwavelength scale from a *single* antenna, which extends the methodology suggested in Ref. [1]. It is observed that the sparse wideband sources can be located in a subwavelength scale from the measurements acquired by only one fixed antenna. Moreover, this imaging device makes no use of any conventional lens, and thus such implementation can find applications in other disciplines, such as, THz, RF or ultrasound imaging. As a by-product, we show that a strongly disordered medium can serves as an efficient apparatus for compressive measurement, which shifts the complexity of devising CS hardware from the design, fabrication and electronic control to a simple and single calibration procedure. Furthermore, unlike most current compressive sensing hardware, this system gives access to many compressive measurements by one single antenna, drastically speeding up acquisition.

Appendix

In this appendix, we will briefly discuss how to derive the Green' function for the disordered medium system. Let
$$E_\theta^0(\mathbf{r},\omega) = \mu_0 \omega^2 G_0(\mathbf{r},\mathbf{r}_\theta;\omega) \quad (A1)$$
be the electrical field associated with an electromagnetic wave at **r** from a source located at $\mathbf{r}_\theta$=(R$_0$cosθ, R$_0$sinθ). As mentioned previously, the Green's function relates the electric field at the position of antenna $\mathbf{r}_d$ to a given point source at $\mathbf{r}_\theta$. To proceed, we employ the coupled-dipole method for solving following coupled electrical integral equations [8], i.e.,

$$E_\theta(\mathbf{r}_d,\omega) = E_\theta^0(\mathbf{r}_d,\omega) + k_0^2 \chi \sum_{j=1}^{J} E(\mathbf{r}_j,\omega) \int_{V_j} G_0(\mathbf{r}_d,\mathbf{r}';\omega) d\mathbf{r}' \quad (A2)$$

and

$$E(\mathbf{r}_i,\omega) = E_\theta^0(\mathbf{r}_i,\omega) + k_0^2 \chi \sum_{j=1}^{J} E(\mathbf{r}_j,\omega) \int_{V_j} G_0(\mathbf{r}_i,\mathbf{r}';\omega) d\mathbf{r}' \quad (A3)$$

$$i=1,2,...,J$$

where $\chi = \varepsilon_r - 1$, $J$ is the total number of cylindrical scatters, and $G_0(\mathbf{r},\mathbf{r}';\omega) = i/4 H_0^{(1)}(k_0|\mathbf{r}-\mathbf{r}'|)$ is the free space Green's function with $H_0^{(1)}$ the first-kind and zero-order Hankel function. Herein we assume that the wavefiled is uniform

over one scatter, which is a good approximation if the scatter is smaller than the operational wavelength λ. $k_0$ is the wavenumber dictated by $k_0 = 2\pi/\lambda$. After solving Eq. (A2) and (A3) for all values of θ and ω, we can immediately obtain the Green's function as $G(\mathbf{r}_d, \mathbf{r}_\theta; \omega) = E_\theta(\mathbf{r}_d, \omega)$, as shown in Fig.2. In our numerical approach, all of the near-field, far-field dipole-dipole interactions and multiple scattering are taken into account rigorously.